\documentclass[aps,prb,reprint,amsfonts,amsmath,amssymb,longbibliography,nofootinbib,floatfix]{revtex4-2}
\usepackage{hyperref}
\usepackage{graphicx}
\usepackage{bm}
\usepackage{newtxtext}
\usepackage[varg]{newtxmath}
\hypersetup{colorlinks=true,linkcolor=blue,citecolor=blue,urlcolor=blue}
\usepackage{mathtools}
\usepackage{mleftright}
\mleftright
\usepackage{orcidlink}

\DeclareMathOperator{\Tr}{Tr}

\DeclareMathOperator{\re}{Re}

\newcommand{\tho}{t_{\mathrm{h}}}
\newcommand{\JK}{J_{\mathrm{K}}}
\newcommand{\Js}{J_{\mathrm{s}}}
\newcommand{\Lth}{\mathcal{L}_{\theta}}
\newcommand{\Whop}{W_{\mathrm{hop}}}
\newcommand{\dth}{\Delta\theta}

\begin{document}

\title{High-harmonic fingerprints of sharp spin twists in a chiral soliton lattice}

\author{\mbox{Atsushi Ono\,\orcidlink{0000-0002-8813-4470}}}
\affiliation{\mbox{Department of Physics, Graduate School of Science, Tohoku University, Sendai 980-8578, Japan}}

\date{\today}

\begin{abstract}
A magnetic field applied perpendicular to the helical axis of a monoaxial chiral helimagnet compresses the spin helix into a coplanar chiral soliton lattice (CSL).
We show that optically driven high-harmonic generation from itinerant electrons coupled to a frozen CSL resolves the lattice-scale structure of the localized twist, rather than the continuum soliton shape.
High-order harmonics remain perturbative for the uniform helix.
As the winding localizes at fixed magnetic period, they grow by many orders of magnitude and acquire a nonperturbative dependence on the drive amplitude.
The growth originates from a spatially nonuniform effective hopping that turns each soliton into a localized dip in the hopping amplitude.
When the degree of winding localization is held fixed, high-order intensities fall by many orders of magnitude as the magnetic period increases toward the continuum limit, where the hopping modulation is spatially smoothed.
High-order harmonics thus resolve a real-space characteristic of the coplanar CSL, even in the absence of scalar chirality and an emergent magnetic field.
\end{abstract}

\maketitle

\section{Introduction}
\label{sec:intro}

Since the first observation of nonperturbative harmonics in ZnO~\cite{Ghimire2011}, high-harmonic generation (HHG) in solids has developed into a strong-field spectroscopic probe of Bloch electrons~\cite{Schubert2014, Vampa2014, Wu2015, Luu2015, Vampa2015Nature, Hohenleutner2015, Tamaya2016, Ndabashimiye2016, TancogneDejean2017, Ikemachi2017, You2017, Ikeda2018, Kruchinin2018, Ghimire2019, Sanari2020, Sanari2020a, Goulielmakis2022}.
Harmonic spectra have been used to reconstruct energy bands~\cite{Vampa2015, Lanin2017, Imai2022} and to probe Berry curvature~\cite{Liu2017, Luu2018, Silva2019, Schmid2021, Yue2023} and Berry phase~\cite{UzanNarovlansky2024}.
The approach now extends well beyond conventional band insulators to graphene and related two-dimensional materials~\cite{Yoshikawa2017, Taucer2017, Hafez2018, Yoshikawa2019, Ikeda2020, Sato2021, Cha2022}, topological insulators and semimetals~\cite{Cheng2020, Kovalev2020, Bai2021, Lv2021, Heide2022, Neufeld2023, Lorenzo2026}, Mott insulators and other correlated systems~\cite{Silva2018, Murakami2018, TancogneDejean2018, Imai2020, Murakami2021, Uchida2022, Murakami2022, Nakano2024, Murakami2025}, and quantum spin systems~\cite{Takayoshi2019, Ikeda2019, Yarmohammadi2024}.

Noncoplanar spin textures can carry scalar chirality and couple to these dynamics through the resulting Berry curvature of the conduction electrons~\cite{Nagaosa2013}.
In a four-sublattice scalar chiral state, reversing the chirality flips the curvature while leaving the bands untouched, and this purely geometric change was shown to modulate the electron--hole trajectories and thereby the longitudinal harmonics~\cite{Ono2024}.
Related fingerprints of static spin textures in HHG have also been identified in two-dimensional magnets, where even harmonics track momentum-space spin texture and Berry curvature~\cite{Jia2020, Gabriele2025}.

A magnetic field applied perpendicular to the helical axis of a monoaxial chiral helimagnet compresses the helix into a chiral soliton lattice (CSL), a periodic array of $2\pi$ twists separated by nearly ferromagnetic segments~\cite{Dzyaloshinskii1964, Kishine2015, Togawa2012, Togawa2016}.
The texture is coplanar, its scalar chirality vanishes identically, and a one-dimensional chain supports no emergent magnetic field, so the Berry-curvature channel of noncoplanar magnets is absent.
In a local spin frame, the texture is encoded entirely in the hopping through an SU(2) matrix on each bond, with the spin-mixing part largest on the bonds at each soliton center.
Itinerant electrons scatter from the soliton array~\cite{Okumura2017}, and a uniform helix is gauge-equivalent to an almost ordinary tight-binding chain, whereas the CSL superlattice opens a ladder of gaps at the folded Brillouin-zone boundary~\cite{Okumura2018}.

In this study, we show that optically driven strong-field HHG can resolve this miniband reconstruction and distinguish the lattice-scale hopping inhomogeneity of a sharp twist from the continuum soliton shape.
We simulate a ferromagnetic Kondo chain coupled to a frozen CSL, comparing textures that share the same $2\pi$ winding per magnetic period $L$ but distribute it differently.
High-order harmonics remain perturbative for the helix and grow by many orders of magnitude as the twist localizes, acquiring a nonperturbative dependence on the drive amplitude; projected models and a magnetic-period scan trace the growth to the lattice-scale modulation of the hopping.

Section~\ref{sec:model} defines the model, the textures, the drive, and the projected models.
Section~\ref{sec:results} presents the results, Sec.~\ref{sec:discussion} discusses their implications, and Sec.~\ref{sec:summary} summarizes.

\section{Model and methods}
\label{sec:model}

\subsection{Kondo chain with a frozen soliton lattice}
\label{sec:hamiltonian}

We consider the one-dimensional ferromagnetic Kondo lattice model, which describes itinerant electrons coupled to classical localized spins.
The Hamiltonian is
\begin{align}
H(t)
&= -\tho \sum_{is}
\mleft[ \mathrm{e}^{-\mathrm{i} A(t)} c^{\dagger}_{i,s} c_{i+1,s} + \mathrm{H.c.} \mright]
\notag \\
&\quad - \JK \sum_{iss'} \bm{S}_{i} \cdot \bm{\sigma}_{ss'} c^{\dagger}_{i,s} c_{i,s'},
\label{eq:model}
\end{align}
where $c_{i,s}^\dagger$ is the creation operator for an electron with spin $s$ at site $i$, $\tho$ is the nearest-neighbor hopping, $\JK>0$ is the Kondo (Hund) coupling, $\bm{S}_{i}$ is a classical spin of unit length, and $\bm{\sigma}$ is the vector of Pauli matrices.
The vector potential $A(t)$ at time $t$ enters through the Peierls substitution.
We set $\tho=\hbar=e=a=1$, with $e$ the elementary charge and $a$ the lattice constant.
At strong coupling, the model reduces to double exchange~\cite{Zener1951, Anderson1955, deGennes1960}; we work at $\JK=2$, where the majority and minority sectors are separated but still hybridized by the spin texture.

The localized spins are frozen during the electron dynamics, i.e., $\dot{\bm{S}}_{i}(t) = 0$, which is an approximation that assumes a separation between the electronic time scale and the much slower collective magnetization dynamics; the same approximation was used for the scalar chiral state in Ref.~\cite{Ono2024}.
The filling is $n_{\mathrm{e}}=0.5$ electrons per site (quarter filling of the spinful chain), corresponding to a half-filled majority sector at $\JK=2$.
Unless stated otherwise, the magnetic period is $L=16$ sites with one $2\pi$ winding per period.

\subsection{Spin textures}
\label{sec:texture}

The coplanar textures $\bm{S}_{i}=(\cos\theta_{i},\sin\theta_{i},0)$, where $\theta_{i}$ is the in-plane angle of $\bm{S}_{i}$, are local minima of the classical energy
\begin{align}
E_{\mathrm{spin}}
= -\sum_{i} \mleft[ \Js \cos\dth_{i} + D \sin\dth_{i} + B_{x}\cos\theta_{i} \mright],
\label{eq:espin}
\end{align}
with $\dth_{i}=\theta_{i+1}-\theta_{i}$ and the fixed winding $\theta_{i+L}=\theta_{i}+2\pi$.
We take the ferromagnetic exchange $\Js=1$ and the Dzyaloshinskii--Moriya interaction $D = \Js \tan(2\pi/L)$, so that the minimizer at transverse field $B_{x}=0$ is the uniform helix, $\dth_{i}=2\pi/L$.
A finite $B_{x}$ concentrates the winding into a soliton; the CSL obtained by continuation in $B_{x}$ from this helix remains locally stable up to $B_{x}\approx 0.5$ for $L=16$, beyond which the full twist collapses onto a single bond.
Each texture can be cyclically centered so that $S_{x}(-x)=S_{x}(x)$ and $S_{y}(-x)=-S_{y}(x)$, with $x$ measured from the soliton center, making explicit the symmetry that enforces the odd-harmonic selection rule.

The degree of localization is measured by
\begin{align}
\Lth
= \frac{L}{(2\pi)^{2}} \sum_{i=1}^{L} (\dth_{i})^{2},
\label{eq:ltheta}
\end{align}
which equals unity for the helix and grows as the winding localizes.
We take $B_{x}=0.4$ as a reference, where $\Lth=2.10$.

Let $c_{i}=(c_{i,\uparrow},c_{i,\downarrow})^{\mathrm{T}}$ be the electron spinor at site $i$, and let $U_{i}$ be a site-dependent SU(2) rotation that aligns the local $z$ axis with $\bm{S}_{i}$.
The substitution $c_{i}=U_{i}\tilde{c}_{i}$ diagonalizes the Kondo term and replaces the spin-independent hopping $-\tho\,\tilde{c}^{\dagger}_{i}\tilde{c}_{i+1}$ by $-\tho\,\tilde{c}^{\dagger}_{i}W_{i}\tilde{c}_{i+1}$, where $W_{i}=U_{i}^{\dagger}U_{i+1}$ is the SU(2) matrix on the bond,
\begin{align}
W_{i}
= \exp\mleft( -\frac{\mathrm{i}}{2}\dth_{i}\sigma_{x} \mright)
= \cos\frac{\dth_{i}}{2}\,\sigma_{0} - \mathrm{i} \sin\frac{\dth_{i}}{2}\,\sigma_{x}.
\label{eq:W}
\end{align}
The spin-conserving and spin-mixing amplitudes are therefore $\tho\cos(\dth_{i}/2)$ and $\tho\sin(\dth_{i}/2)$, respectively.
In the helix, every bond carries the same twist $\dth_{i}=2\pi/L$; in the CSL, the same total winding is concentrated on the soliton bonds.

\subsection{Driving and observables}
\label{sec:protocol}

The chain is driven by a continuous wave with a Gaussian ramp,
\begin{align}
A(t)
= -\frac{F_{0}}{\varOmega}\sin(\varOmega t)
\times
\begin{cases}
\mathrm{e}^{-t^{2}/(2\tau^{2})} & (t<0),\\
\hfil 1 & (t\geq 0),
\end{cases}
\label{eq:field}
\end{align}
where $F(t)=-\partial_{t}A(t)$ is the electric field and $F_{0}$ is its steady-state amplitude.
This drive, with $\varOmega=2\pi/50$ and $\tau=6$, follows Ref.~\cite{Ono2024}.
The relevant dimensionless amplitude is the ratio of the vector-potential amplitude $A_{0}=F_{0}/\varOmega$ to the magnetic reciprocal-lattice vector $Q=2\pi/L$,
\begin{align}
\alpha
= \frac{A_{0}}{Q}
= \frac{F_{0}L}{2\pi\varOmega}.
\label{eq:alpha}
\end{align}
At the reference point $\alpha=1$ ($F_{0}\approx 0.0494$ for $L=16$), the vector-potential amplitude equals one magnetic reciprocal-lattice vector, $A_{0}=Q$.
Unless an amplitude scan is shown, all spectra are at this reference amplitude.

Folding the undriven Hamiltonian of Eq.~\eqref{eq:model} into one magnetic cell of $L$ sites gives the $2L\times2L$ Bloch Hamiltonian $h_{k}$ in the site--spin basis,
\begin{align}
(h_{k})_{i,i+1}
= -\tho\,\mathrm{e}^{\mathrm{i} k}\sigma_{0},
\quad
(h_{k})_{i,i}
= -\JK\,\bm{S}_{i}\cdot\bm{\sigma},
\label{eq:bloch}
\end{align}
where $i$ labels sites in the magnetic unit cell, with $i=L$ understood to wrap to $i=1$, and $(h_{k})_{i+1,i}=(h_{k})_{i,i+1}^{\dagger}$.
We place the Bloch phase on the absolute lattice coordinate, $c_{mL+i}=N_{k}^{-1/2}\sum_{k}\mathrm{e}^{\mathrm{i} k(mL+i)}c_{k,i}$, where $m$ labels magnetic cells and $N_{k}$ is their number, so that every nearest-neighbor bond, including the wrap-around bond from $i=L$ to $i=1$, carries the factor $\mathrm{e}^{\mathrm{i} k}$.
The Peierls substitution in Eq.~\eqref{eq:model} is the replacement $k\to k-A(t)$ in $h_{k}$.

The one-particle density matrix $\rho_{k}$ in the magnetic Bloch basis obeys the von Neumann equation with a relaxation-time term,
\begin{align}
\frac{\mathrm{d}\rho_{k}}{\mathrm{d}t}
= -\mathrm{i} \mleft[ h_{k-A(t)}, \rho_{k} \mright]
- \varGamma \mleft( \rho_{k} - \rho_{k}^{(0)} \mright),
\label{eq:eom}
\end{align}
where $\rho_{k}^{(0)}$, the undriven Fermi sea of $h_{k}$ at the global chemical potential $\mu$, is held fixed during the drive, and the relaxation rate is set to $\varGamma=0.1$ unless stated otherwise.
The electric current is
\begin{align}
J(t)
= \frac{1}{LN_{k}} \sum_{k} \Tr \mleft[ \rho_{k}(t)\, \partial_{q}h_{q}\big|_{q=k-A(t)} \mright],
\label{eq:current}
\end{align}
where the sum is over the $N_{k}$ crystal momenta in the magnetic Brillouin zone.
The harmonic spectrum is
\begin{align}
I(\omega) = \omega^{2} \mleft| J(\omega) \mright|^{2},
\quad
J(\omega) = \frac{1}{\Delta T}\int_{t_{1}}^{t_{2}} \mathrm{d}t\, J(t)\, \mathrm{e}^{\mathrm{i}\omega (t-t_{1})},
\label{eq:intensity}
\end{align}
with $t_{1}=500$, $t_{2}=1000$, and $\Delta T=t_{2}-t_{1}$.
The window covers ten optical cycles in the periodic steady state after the ramp and places the harmonics exactly on the discrete frequencies $n\varOmega$; we write $I_{n}=I(n\varOmega)$.
Calculations use $N_{k}=64$ momenta and a fourth-order Runge--Kutta integrator with time step $0.01$ starting at $t=-50$.
At the $B_{x}=0.4$ reference, raising $N_{k}$ to $256$ changes $I_{11}$ by $2\%$; the harmonics are insensitive to halving the time step and to using five instead of ten optical cycles in Eq.~\eqref{eq:intensity}.
Intensities below ${\sim}10^{-32}$ are numerical noise.

\subsection{Projected models and decompositions}
\label{sec:projected}

We compare Eq.~\eqref{eq:model} with spinless chains obtained by projecting the transformed hopping onto the local majority spinor.
These chains obey the same von Neumann equation \eqref{eq:eom}, with $h_{k}$ the $L\times L$ Bloch Hamiltonian defined by the hoppings specified below.
In the local frame, the spin-conserving matrix element of Eq.~\eqref{eq:W} is real, $\cos(\dth_{i}/2)$.
A U(1) gauge choice for the majority spinor places a phase $\dth_{i}/2$ on each bond, giving the adiabatic hopping
\begin{align}
t_{i}^{\mathrm{eff}}
= \tho \cos\frac{\dth_{i}}{2}\, \mathrm{e}^{\mathrm{i}\dth_{i}/2},
\quad
|t_{i}^{\mathrm{eff}}| = \tho \mleft| \cos\frac{\dth_{i}}{2} \mright|.
\label{eq:teff}
\end{align}
For the distributed textures used below, for which $\cos(\dth_{i}/2)>0$, the bond phases in Eq.~\eqref{eq:teff} accumulate to the holonomy $\prod_{i}\mathrm{e}^{\mathrm{i}\dth_{i}/2}=-1$ of one $2\pi$ winding.
We call this \emph{modulated hopping}.
In one dimension, those phases can be gauged to a single boundary phase of $\pi$ without changing the Brillouin-zone-integrated current.
\emph{Uniform hopping} replaces every amplitude by its cell average, removing the spatial modulation of $|t_{i}^{\mathrm{eff}}|$ while leaving the holonomy unchanged (the boundary phase of $\pi$ is kept).
For the helix, the amplitudes are already uniform and the two hoppings coincide.

We also consider two further decompositions.
Let $V_{k}(t)$ be the matrix of instantaneous eigenvectors of the modulated-hopping Hamiltonian $h_{k-A(t)}$.
With $\tilde\rho_{k}=V_{k}^{\dagger}\rho_{k}V_{k}$ and $\tilde v_{k}=V_{k}^{\dagger} [\partial_{q}h_{q}\big|_{q=k-A(t)} ] V_{k}$, the current splits into the band-diagonal (intraband) and off-diagonal (interband) components
\begin{align}
J_{\mathrm{intra}}(t)
&= \frac{1}{LN_{k}}\sum_{k,n}\tilde\rho_{nn}(k,t)\,\tilde v_{nn}(k,t),
\label{eq:intra}
\\
J_{\mathrm{inter}}(t)
&= \frac{1}{LN_{k}}\sum_{k,n\neq m}\tilde\rho_{nm}(k,t)\,\tilde v_{mn}(k,t),
\label{eq:inter}
\end{align}
which sum to $J(t)$ in the time domain.
The corresponding intensities $I_{n}^{\mathrm{intra}}$ and $I_{n}^{\mathrm{inter}}$ follow by substituting these currents for $J$ in Eq.~\eqref{eq:intensity}.
On the projected chain, the current is resolved into bond contributions
\begin{align}
J(t)
&= \frac{1}{L}\sum_{i=1}^{L} j_{i}(t),
\label{eq:bond}
\end{align}
where
\begin{align}
j_{i}(t)
&= \frac{2}{N_{k}}\sum_{k} \re\mleft[\rho_{i,i+1}(k,t)\,(\partial_{q}h_{q})_{i+1,i}\big|_{q=k-A(t)}\mright],
\label{eq:bondj}
\end{align}
with $i=L$ understood to wrap to $i=1$.
The complex amplitudes $j_{i}(n\varOmega)$ are extracted with the same Fourier window as in Eq.~\eqref{eq:intensity}.

\section{Results}
\label{sec:results}

\begin{figure}[t]\centering
\includegraphics[scale=1]{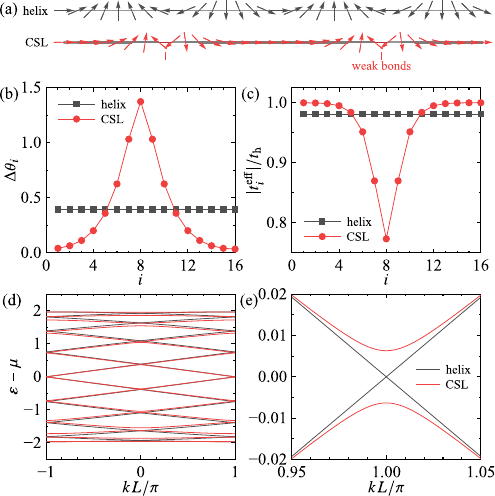}
\caption{Static structure of the CSL ($L=16$; CSL at $B_{x}=0.4$).
(a)~In-plane spins of the uniform helix (gray) and the CSL (red) over two magnetic periods.
(b)~Bond twist $\dth_{i}$ over one magnetic period.
(c)~Projected amplitude $|t_{i}^{\mathrm{eff}}|/\tho$ of Eq.~\eqref{eq:teff}.
(d)~Majority-spin minibands of modulated hopping in the magnetic Brillouin zone, for the uniform helix (gray) and the CSL (red).
(e)~Enlargement of the magnetic Brillouin-zone edge in (d) for $\varepsilon-\mu\in[-0.02,0.02]$.}
\label{fig:weaklink}
\end{figure}

Figure~\ref{fig:weaklink} shows the equilibrium electronic structure of the CSL.
At $B_{x}=0.4$ [Fig.~\ref{fig:weaklink}(a)], the bond twist reaches $\dth_{\max}=1.37$ at the soliton center [Fig.~\ref{fig:weaklink}(b)], so the projected amplitude of Eq.~\eqref{eq:teff} decreases to $0.773 \tho$ there, while the bonds between solitons stay nearly uniform [Fig.~\ref{fig:weaklink}(c)].
To the electrons, a CSL is an almost homogeneous chain interrupted by periodically repeated weaker bonds of magnetic origin.
The helix distributes the same winding equally, $\dth_{i}=2\pi/16\approx0.39$, and has no spatial amplitude modulation.

The reciprocal-space counterpart of these weaker bonds is the reconstructed majority-spin miniband structure of Fig.~\ref{fig:weaklink}(d).
Writing $|t_{i}^{\mathrm{eff}}|=\overline{|t^{\mathrm{eff}}|}+\delta t_{i}$, with the overline a cell average, the Fourier components of $\delta t_{i}$ couple momenta $k$ and $k+q$ and open avoided crossings between the folded bands, which are absent for the helix.
The miniband reconstruction is the Bloch representation of the periodic hopping modulation.
At $n_{\mathrm{e}}=0.5$, the majority sector is half filled; with the $\pi$ holonomy of one $2\pi$ winding per cell, the Fermi points fold to the magnetic Brillouin-zone boundary.
Figure~\ref{fig:weaklink}(e) shows that the helix bands cross at the chemical potential $\mu$, whereas the CSL opens a small gap there ($\approx 0.013$), far below the bandwidth and $\varGamma=0.1$.
Although this gap would render the projected undriven CSL a clean zero-temperature band insulator, it is unresolved on the relaxation scale $\varGamma=0.1$, and the driven fundamental remains Drude-like under the present driving conditions.

\begin{figure*}[t]\centering
\includegraphics[scale=1]{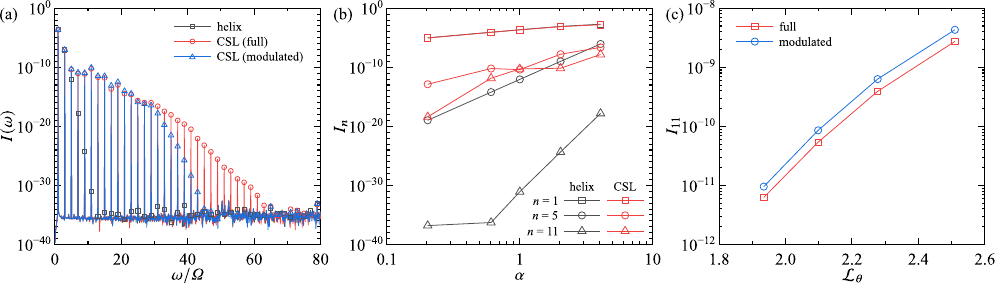}
\caption{Emergence of high harmonics.
(a)~Harmonic spectra at $\alpha=1$ of the full model for the helix (gray) and the CSL at $B_{x}=0.4$ (red), and of modulated hopping for the same CSL (blue).
(b)~$I_{1}$, $I_{5}$, and $I_{11}$ versus $\alpha$ for the helix (gray) and the CSL (red) at $B_{x}=0.4$; squares, circles, and triangles mark $n=1$, $5$, and $11$.
(c)~$I_{11}$ versus the localization measure $\Lth$ of Eq.~\eqref{eq:ltheta} along the $B_{x}$ continuation at $\alpha=1$, for $B_{x}=0.35$--$0.5$, shown for the full model (red) and modulated hopping (blue).}
\label{fig:emergence}
\end{figure*}

Under the drive, the harmonic spectra nevertheless separate the CSL from the helix by many orders of magnitude [Fig.~\ref{fig:emergence}(a)].
The fundamental harmonic is a Drude-like response and is insensitive to the texture, $I_{1}\approx 2.2\times10^{-4}$ in both cases.
For the helix, which is gauge-equivalent to an almost uniform tight-binding metal, the spectrum is that of a one-dimensional metal: a Drude fundamental and a steep perturbative tail ($I_{5}=9.0\times10^{-13}$, converged with respect to $N_{k}$) that reaches $I_{11}=8.5\times10^{-32}$ at the edge of the numerical floor.
The CSL instead develops a high-order shoulder, which sits near $n=11$ for this $\varOmega$, with $I_{11}=5.4\times10^{-11}$, twenty orders of magnitude above the helix.

The field-amplitude scan in Fig.~\ref{fig:emergence}(b) confirms the same contrast without relying on the spectral shape.
For the helix, $I_{1}$ and $I_{5}$ follow the perturbative laws $\propto\alpha^{2}$ and $\propto\alpha^{10}$, respectively, from the weakest drive, and $I_{11}$ follows $\propto\alpha^{22}$ for $\alpha\gtrsim 1$.
The CSL fundamental remains Drude-like, but already at $\alpha=0.20$ the fifth harmonic exceeds the helix value by six orders of magnitude, and $I_{11}$ for the CSL rises far more slowly than the perturbative $\alpha^{22}$ scaling.
Between $\alpha=1$ and $2$, the CSL intensity $I_{11}$ changes only from $5.4\times10^{-11}$ to $6.9\times10^{-11}$, whereas a perturbative eleventh harmonic would grow by a factor $2^{22}\approx 4\times10^{6}$.
Even at $\alpha=4$, $I_{11}$ for the helix remains many orders below the CSL ($I_{11}=1.5\times10^{-18}$).
The high-order CSL response is therefore nonperturbative in both its field-amplitude scaling and its spectral signature.

Figure~\ref{fig:emergence}(c) tracks $I_{11}$ against the localization measure $\Lth$ of Eq.~\eqref{eq:ltheta} along the $B_{x}$ continuation at fixed $L=16$ and $\alpha=1$.
The scan starts at $B_{x}=0.35$, below which the Fermi-level avoided crossing opened by the $2k_{\mathrm{F}}$ Fourier component of $\delta t_{i}$ is not resolved on the magnetic Brillouin-zone mesh $\delta k=2\pi/(L N_{k})$, where $k_{\mathrm{F}}$ is the Fermi wavevector of the half-filled majority chain.
The reference texture of panels (a) and (b), $B_{x}=0.4$ ($\Lth=2.10$), lies on this curve at $I_{11}=5.4\times10^{-11}$.
The intensity grows from $6.3\times10^{-12}$ at $B_{x}=0.35$ ($\Lth=1.93$) to $2.7\times10^{-9}$ at $B_{x}=0.5$ ($\Lth=2.51$), the highest field at which the $L=16$ lattice CSL remains a distributed $2\pi$ twist.
Beyond $B_{x}\approx 0.5$, the winding collapses onto a single bond and $\Lth$ reaches the single-bond value $\Lth=L$ of this monotonic-winding family.
At $B_{x}=0.5$, the zone-edge gap has grown to $\approx 0.048$ and contributes to $I_{11}$.
Low orders are far less sensitive; $I_{5}$ stays near its helix value up to $B_{x}\approx 0.2$.

Modulated hopping of Eq.~\eqref{eq:teff} reproduces the full spectra closely: the ratio of full-model to modulated-hopping $I_{n}$ is $1.01$, $0.84$, and $0.63$ at $n=1$, $5$, and $11$, respectively.
The enhancement therefore survives the adiabatic projection, and nonadiabatic spin mistracking is a modest correction, consistent with $\max_{t}|J(t)|\,\dth_{\max}/(2\JK n_{\mathrm{e}})\approx 0.16$ during the drive, which compares the current-based velocity $|J|/n_{\mathrm{e}}$ times the peak bond twist to the exchange splitting $2\JK$.

\begin{figure}[t]\centering
\includegraphics[scale=1]{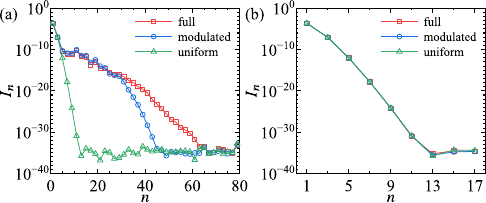}
\caption{Modulated versus uniform hopping ($\alpha=1$).
(a)~Odd-harmonic intensities $I_{n}$ for the CSL at $B_{x}=0.4$: full model, modulated hopping, and uniform hopping.
(b)~The same for the helix, shown for $n\le 17$.}
\label{fig:gauges}
\end{figure}

Figure~\ref{fig:gauges} separates the two contributions in Eq.~\eqref{eq:teff}.
Helix and CSL carry the same holonomy.
Uniform hopping keeps that holonomy but averages out the amplitudes, and the high orders collapse to the helix floor; $I_{11}$ drops from $8.6\times10^{-11}$ for modulated hopping to $1.2\times10^{-31}$, and the uniform-to-modulated ratio is $0.018$ already at $n=5$.
The high-harmonic enhancement requires the spatial modulation of $|t_{i}^{\mathrm{eff}}|$ (i.e., the weak bonds of Fig.~\ref{fig:weaklink}), whereas the shared $\pi$ holonomy alone does not produce it.
The full model and modulated hopping remain close through the high-order shoulder of Fig.~\ref{fig:gauges}(a).
They separate beyond $n\approx 31$, where $n\varOmega$ approaches the majority--minority splitting $2\JK$.
The adiabatic projection retains only the majority sector, so the additional full-model weight is a minority-sector response omitted by Eq.~\eqref{eq:teff}.

\begin{figure}[t]\centering
\includegraphics[scale=1]{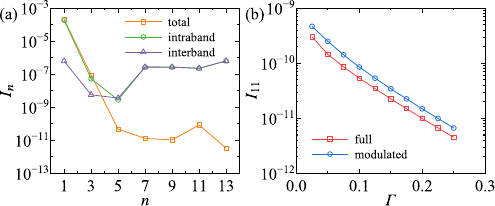}
\caption{Dynamical character of the high-order response ($B_{x}=0.4$, $\alpha=1$).
(a)~Decomposition of the modulated-hopping current into intra- and interband pieces for odd $n\le 13$.
(b)~$I_{11}$ versus $\varGamma$ for the full CSL and modulated hopping.}
\label{fig:dynamics}
\end{figure}

That high-order shoulder is the residual left by cancellation between the intra- and interband currents in Eqs.~\eqref{eq:intra} and \eqref{eq:inter} [Fig.~\ref{fig:dynamics}(a)].
At $n=1$, the current is overwhelmingly intraband, $I_{1}^{\mathrm{inter}}/I_{1}^{\mathrm{intra}}=0.003$, as expected for a Drude fundamental.
For $n\ge 5$ both pieces exceed the total and nearly coincide; at $n=11$, $I_{11}^{\mathrm{intra}}=2.30\times10^{-7}$ and $I_{11}^{\mathrm{inter}}=2.32\times10^{-7}$, each three orders of magnitude above the total $I_{11}=8.6\times10^{-11}$.
The intensities are those of $J_{\mathrm{intra}}$ and $J_{\mathrm{inter}}$ separately, so $J_{\mathrm{intra}}(n\varOmega)\approx -J_{\mathrm{inter}}(n\varOmega)$ makes the two intensities coincide while the residual $J(n\varOmega)$ stays small.
This cancellation is analogous to the intra- and interband interference discussed for conventional solids in Ref.~\cite{Wang2018}.
The cancellation is sensitive to relaxation: as $\varGamma$ grows from $0.05$ to $0.2$, $I_{11}$ of the full model in Fig.~\ref{fig:dynamics}(b) falls from $1.5\times10^{-10}$ to $1.0\times10^{-11}$ while $I_{1}$ changes little.
The full model and modulated hopping remain close at every $\varGamma$; the helix and the uniform-hopping model remain at the numerical floor.

\begin{figure}[t]\centering
\includegraphics[scale=1]{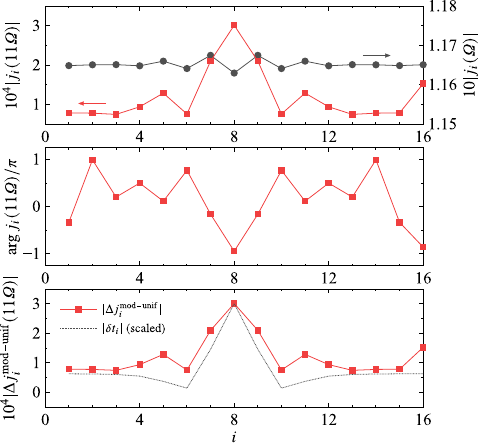}
\caption{Bond-resolved harmonic response ($B_{x}=0.4$, $\alpha=1$; one magnetic period).
Top:~$|j_{i}(11\varOmega)|$ of modulated hopping (left axis) and $|j_{i}(\varOmega)|$ (right axis).
Middle:~Phase of $j_{i}(11\varOmega)$.
Bottom:~$|\Delta j_{i}(11\varOmega)|$ between modulated and uniform hopping, and $|\delta t_{i}|$ (rescaled).}
\label{fig:bond}
\end{figure}

The bond decomposition of Eqs.~\eqref{eq:bond} and \eqref{eq:bondj} locates the large local contributions that underlie this residual (Fig.~\ref{fig:bond}).
The top panel of Fig.~\ref{fig:bond} shows the eleventh-harmonic amplitude $|j_{i}(11\varOmega)|$ of modulated hopping, which peaks at the soliton-center bond, where $\dth_{i}$ and $|\delta t_{i}|$ are largest, with a secondary peak on the bond that closes the magnetic cell and a weak background elsewhere.
The fundamental is spatially uniform, $|j_{i}(\varOmega)|\approx0.116$ on every bond.
In the middle panel, the phases of $j_{i}(11\varOmega)$ vary strongly across the cell, and the spatial sum nearly cancels, $|\sum_{i}j_{i}(11\varOmega)|/\sum_{i}|j_{i}(11\varOmega)|=0.055$, whereas the fundamental adds fully in phase (the same ratio is $1.00$).
A strong local eleventh-harmonic response at the solitons is therefore compatible with a much smaller emitted intensity $I_{11}$: the observed harmonic is the residual of a spatial cancellation.
The bottom panel compares the difference between modulated and uniform hopping with the hopping modulation $|\delta t_{i}|$; the Pearson correlation coefficient of the two profiles over one magnetic cell is $0.89$, so the modulation-induced change in the local response spatially tracks the hopping modulation.
In a wave-packet picture, carriers driven by $A(t)$ are coherently reflected and transmitted at the periodically repeated sharp-twist regions, and the high harmonics are the radiation emitted by this driven coherent transport through the weak bonds.

\begin{figure*}[tb]\centering
\includegraphics[scale=1]{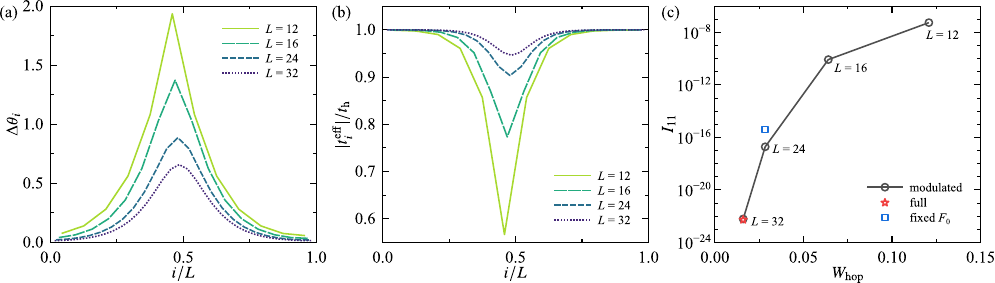}
\caption{Magnetic-period scan at matched localization $\Lth\approx 2.10$ (modulated hopping).
(a)~Bond twist $\dth_{i}$ and (b)~projected amplitude $|t_{i}^{\mathrm{eff}}|/\tho$ versus $i/L$ for $L=12$, $16$, $24$, and $32$ ($B_{x}=0.64$, $0.40$, $0.185$, and $0.105$); the wrap-around bond is repeated at both cell edges.
(c)~$I_{11}$ versus $\Whop$ of Eq.~\eqref{eq:whop}.
Circles are at $\alpha=1$; the star is the full model at $L=32$; the square is the same $L=24$ texture driven at the $F_{0}$ of the $L=16$ reference ($\alpha=1.5$).}
\label{fig:period}
\end{figure*}

Because the enhanced local high-order response is concentrated around those weak bonds, the lattice-scale sharpness of the twist can affect the harmonics even when the continuum soliton shape does not change.
For $L=12$, $16$, $24$, and $32$, we retune $D=\Js\tan(2\pi/L)$ and choose $B_{x}=0.64$, $0.40$, $0.185$, and $0.105$ so that $\Lth\approx 2.10$ in all cases.
The resulting profiles are self-similar, $\theta_{i}\approx \varTheta(i/L)$ with $L\,\dth_{\max}\approx 21$--$23$ [Fig.~\ref{fig:period}(a)], so the bond twist scales as $L^{-1}$.
Consequently, the amplitude modulation
\begin{align}
\Whop
= \mleft[ \frac{1}{L}\sum_{i}(\delta t_{i})^{2} \mright]^{1/2}
\label{eq:whop}
\end{align}
scales as $L^{-2}$, with $L^{2}\Whop\approx 16.4$ for $L\geq 16$ and $17.4$ at $L=12$ [Figs.~\ref{fig:period}(b) and \ref{fig:period}(c)].

At matched $\Lth$, $I_{11}$ still falls by many orders of magnitude as $L$ increases [Fig.~\ref{fig:period}(c)], from $5.6\times10^{-8}$ at $L=12$ to $6.2\times10^{-23}$ at $L=32$.
The $L=16$ reference and $L=32$ share the same localization yet differ by twelve orders of magnitude in $I_{11}$.
Because $\alpha=1$ is held along this scan, Eq.~\eqref{eq:alpha} implies $F_{0}\propto L^{-1}$, so the drive amplitude varies together with $\Whop$.
That covariation does not account for the full drop: the same $L=24$ texture, driven at the $F_{0}$ of the $L=16$ reference ($\alpha=1.5$), remains five orders of magnitude below the reference.
The fundamental remains Drude-like, of order $10^{-5}$--$10^{-4}$, for every $L$.
The full-model $I_{11}$ at $L=32$ agrees with the modulated-hopping result.
High-order harmonics are therefore strongly sensitive to the lattice-scale sharpness of the twist, even when the continuum soliton shape is held fixed.

\section{Discussion}
\label{sec:discussion}

A sharp spin twist produces large bond twists $\dth_{i}$.
These modulate the adiabatic hopping amplitude and create a periodic spatial modulation of $|t_{i}^{\mathrm{eff}}|$.
The Bloch counterpart of this periodic bond modulation is a reconstructed magnetic miniband structure.
The strong field drives quantum dynamics within these minibands, and the resulting nonlinear current generates the high harmonics.
The miniband reconstruction is an equilibrium property of the itinerant CSL~\cite{Okumura2017, Okumura2018}.
Optically driven HHG probes the lattice-scale structure of that reconstruction.

The channel identified here complements the Berry-curvature channel of the scalar chiral state~\cite{Ono2024}.
In the noncoplanar case, the texture deflects electron--hole trajectories through the anomalous velocity; in the coplanar CSL, it modulates the magnitude of the projected hopping.
Both depend on the relation between the magnetic and electronic length scales, but they act through different couplings.
The coplanar HHG channel is also distinct from the quadratic optical response of a conical chiral magnet~\cite{Okumura2021}, from circular dichroism in third-harmonic generation associated with vector spin chirality~\cite{Huang2026}, and from harmonics pumped by a precessing magnetic order~\cite{Ly2022, Ly2023, Ly2025, Ly2025a, Ly2026}.
It is also distinct from harmonics of the magnetization driven by an ac magnetic field in the CSL~\cite{Tsuruta2016, Clements2018, Kishine2020}: those signals track slow spin motion, whereas here the texture is held frozen and the radiation comes from the charge current of the itinerant electrons.

The self-similar family of Fig.~\ref{fig:period} varies the lattice-scale sharpness at fixed continuum shape.
The dilute-soliton regime of the continuum CSL, with the local twist held fixed and only the spacing varied~\cite{Kishine2015, Masaki2018, Masaki2020}, is left for separate work.
There $\Whop\propto L^{-1/2}$ because $\Whop^{2}$ receives a contribution from one kink per cell, in contrast to the self-similar family of Fig.~\ref{fig:period}, for which $\Whop\propto L^{-2}$.
Even then the far-field intensity need not scale simply with soliton density $n_{\mathrm{sol}}$: a coherent superposition of identical kink amplitudes would give $J_{n}\propto n_{\mathrm{sol}}$ and $I_{n}\propto n_{\mathrm{sol}}^{2}$, whereas the bond-resolved cancellation $|\sum_{i}j_{i}|/\sum_{i}|j_{i}|\approx 0.055$ already shows that spatial interference is severe.

Monoaxial chiral helimagnets such as CrNb$_3$S$_6$ provide metallic hosts in which a transverse field tunes the texture continuously from a helix to a CSL~\cite{Togawa2012, Togawa2013, Togawa2016}.
Along the usual field sweep, the soliton spacing $L$ grows toward the isolated-kink limit while the core width stays of order the helical pitch.
This corresponds to the dilute-soliton regime discussed above.
The hopping modulation appears as the helix deforms into a CSL, and the cell-averaged modulation strength then decreases as the soliton density $n_{\mathrm{sol}}=1/L$ decreases, so a non-monotonic high-order intensity is possible.
The CSL period already controls dc magnetotransport through electron scattering from the soliton array~\cite{Togawa2013, Okumura2017}; the present results indicate that, when the electrons are driven at optical frequencies, the same spatial modulation of hopping is reflected in the high harmonics.

In our units, the drive $\varOmega=2\pi/50$ lies far below the bandwidth, in the multiterahertz range for typical hoppings, and $\alpha=1$ corresponds to a vector-potential excursion of one magnetic reciprocal-lattice vector.
At fixed commensurate period $L$, increasing the field localizes each $2\pi$ twist, and high-order harmonics are predicted to grow strongly and to depend sensitively on the relaxation rate, remaining far above the helix intensities and departing from the perturbative $I_{n}\propto\alpha^{2n}$ law of the helix, while the fundamental remains essentially unchanged and lower harmonics are much less sensitive than the high-order response.
Joint measurements of harmonic order, drive amplitude, frequency, and relative harmonic phase may later disentangle $L$ from the local twist profile.

Natural extensions include fillings with $\mu$ inside a large miniband gap, for which an interband recombination picture may become applicable, and unfrozen spins, which couple the hopping modulation to the collective CSL magnon modes~\cite{Shimamoto2022}.
A two-dimensional counterpart would be a skyrmion crystal of finite magnetic period, which combines the hopping-modulation channel of the CSL with the Berry-curvature channel of the scalar chiral state~\cite{Nagaosa2013, Ono2024}, so that the effective magnetic period is the skyrmion spacing along the drive.

\section{Summary}
\label{sec:summary}

A coplanar chiral soliton lattice, though free of scalar chirality and an emergent magnetic field, leaves a strong imprint on optically driven high-harmonic generation.
High-order harmonic intensities grow by many orders of magnitude as the spin twist localizes.
This growth is generated by quantum dynamics within the minibands that arise from the periodic weak-bond modulation created by the solitons.
Along the self-similar CSL family at fixed localization, the hopping inhomogeneity scales as $L^{-2}$ toward the continuum limit, while the high-order intensities fall by many orders of magnitude.
High-order HHG is thereby a spectroscopic fingerprint of the lattice-scale structure of sharp spin twists in chiral magnets.

\begin{acknowledgments}
This work was supported by JSPS KAKENHI Grants No.\ JP23K13052, No.\ JP24K00563, No.\ JP26K06993, and No.\ JP26K00646.
The numerical calculations were performed using the facilities of the Supercomputer Center, the Institute for Solid State Physics, the University of Tokyo.
\end{acknowledgments}

\bibliography{reference}

\end{document}